\documentclass[aps,prb,twocolumn]{revtex4}

\usepackage{amsmath}
\usepackage{amssymb}
\usepackage{color}
\usepackage{xcolor}
\usepackage{tikz}
\usepackage{bbm}
\usepackage{hyperref}
\usepackage{ulem}
\renewcommand{\k}{\mathbf{k}}

\begin{document}
\title{Honeycomb rare-earth magnets with anisotropic exchange interactions}
\author{Zhu-Xi Luo$^{1,2}$}
\author{Gang Chen$^{3,4}$}
\thanks{gangchen.physics@gmail.com}
\affiliation{$^{1}$Kavli Institute for Theoretical Physics, University of California, Santa Barbara, CA 93106}
\affiliation{$^{2}$Department of Physics and Astronomy, University of Utah, Salt Lake City, UT 84102 }
\affiliation{$^{3}$Department of Physics and HKU-UCAS Joint Institute for Theoretical and Computational Physics at Hong Kong, The University of Hong Kong, Hong Kong, China}
\affiliation{$^{4}$State Key Laboratory of Surface Physics and Department of Physics, Fudan University, Shanghai 200433, China}
\date{\today}

\begin{abstract}
We study the rare-earth magnets on a honeycomb lattice, and are particularly interested in the experimental consequences of the highly anisotropic spin interaction due to the spin-orbit entanglement. We perform a high-temperature series expansion using a generic nearest-neighbor Hamiltonian with anisotropic interactions, and obtain the heat capacity, the parallel and perpendicular spin susceptibilities, and the magnetic torque coefficients. We further examine the electron spin resonance linewidth as an important signature of the anisotropic spin interactions. Due to the small interaction energy scale of the rare-earth moments, it is experimentally feasible to realize the strong-field regime. Therefore, we perform the spin-wave analysis and study the possibility of topological magnons when a strong field is applied to the system. The application and relevance to the rare-earth Kitaev materials are discussed.
\end{abstract}
\maketitle

\section{Introduction}

Spin liquid candidates are often being searched among geometrically frustrated systems, such as triangular \cite{Anderson}, kagom\'{e} \cite{Sachdev} or pyrochlore \cite{Anderson1956} lattices. 
This is quite reasonable as the geometrical frustration could
lead to a large number of degenerate or nearly-degenerate classical ground states for commonly studied Heisenberg models and thus enhance quantum fluctuations
when the quantum effects are included. 
However, the destabilization of simple magnetically ordered states and driving a disordered one can happen even on unfrustrated lattices, by exploiting the power of anisotropic interactions \cite{Krempa}; the Kitaev honeycomb model \cite{Kitaev} is a representative example of the latter. Besides being an academic interest, anisotropic spin interactions are also inevitable in realistic magnetic materials, especially those with heavy atoms. A large number of spin liquid candidates are known experimentally to possess a significant spin-orbit coupling, leading to rather anisotropic spin interactions~\cite{PhysRevX.1.021002,Pan,Gingras,YbMgGa1,YbMgGa2,RuCl3}. 
Beyond the current interest in the spin liquid physics, understanding the 
relationship between the magnetic properties and the anisotropic
spin interactions is a frontier topic in the field of quantum magnetism.

The most commonly studied anisotropic magnets on unfrustrated lattices are the $4d/5d$ magnets \cite{RuCl3,Iridates1}
that include the honeycomb iridates and RuCl$_3$. 
Due to the possible proximity to Kitaev physics, 
these materials were referred as Kitaev materials. 
Due to the spatial extension of the $4d/5d$ electron wavefunctions, 
the exchange interactions between the local moments 
are usually beyond the nearest neighbors. 
Moreover, the iridates often suffer from a strong neutron absorption such that the data-rich neutron scattering measurement can be difficult. 
In comparison, the rare-earth family has the advantages of much stronger spin-orbit couplings and much more localized $4f$ orbitals \cite{PhysRevB.95.085132,Jang,RauYb}, and the exchange interactions often restrict to first neighbors. This makes the understanding of the modeling Hamiltonian more accessible. 
In addition, the rare-earth magnets do not have 
the neutron absorption issue that prevails in iridates. 
Furthermore, their smaller energy scales allow for the 
possibility to quantitatively understand their Hamiltonian 
through the external magnetic fields. %~\cite{Ni} 
However, rare-earth materials have only been well-investigated 
on frustrated lattices \cite{Pyrochlore1, YbMgGa1}.

In this paper, we will study the rare-earth magnets on the  
unfrustrated honeycomb lattice, and pursue an understanding 
of the experimental consequence of the spin-orbital entanglement
on the honeycomb structure. We start by exploring the thermodynamic
properties of a generic model with the nearest neighbor interactions. 
It is well-known that the anisotropic exchange couplings could appear in the temperature dependence of the 
thermodynamic quantities such as the specific heat, 
spin susceptibility \cite{PhysRevB.96.144414} and magnetotropic coefficients \cite{Torque,PhysRevLett.122.197202}. 
Especially for the spin susceptibility and magnetic 
torque, magnetic fields along different directions 
induce magnetization of different magnitudes, leading to the anisotropic spin susceptibility \cite{PhysRevB.82.064412, PhysRevB.91.094422,PhysRevB.91.144420, PhysRevB.91.180401, freund2016single, PhysRevB.98.100403}
and the angular dependence of the 
magnetic torque \cite{PhysRevLett.118.187203,PhysRevB.98.205110,modic2018resonant,PhysRevB.99.081101}, and providing a natural detection of the 
intrinsic spin anisotropy in the system. 
To go beyond the thermodynamic properties, we further consider the electron spin resonance (ESR) measurement \cite{Oshikawa,PhysRevB.96.241107} of the system. 
The ESR measurement turns out to be a very sensitive probe of the magnetic anisotropy and is especially useful for the study of the strong spin-orbit-coupled quantum materials, and we compute the ESR linewidth to reveal the intrinsic spin anisotropy of the spin interactions. 

Due to the small energy scale of the interaction 
between the rare-earth local moment,
it is ready to apply a small magnetic field in the laboratory to change the magnetic
state into a fully polarized one. For such a simple product state, the magnetic excitation can be readily worked out from the linear spin wave theory. We further consider the spin wave spectrum and explore the possibility of topological magnons \cite{Shindou1, Shindou2, Matsumoto, Matsumoto1, Owerre1, Owerre2, LChen,McClarty}. We find the magnon spectrum supports non-trivial topological band structure. This feature can be manifested in thermal Hall transport measurements.

The remaining parts of the paper are organized as follows. 
In Sec.~\ref{sec:model}, we introduce the nearest-neighbor 
spin Hamiltonian, followed by the high-temperature 
analysis of heat capacity, spin susceptibilities 
and magnetic torque coefficient in 
Sec~\ref{sec:thermo}. 
Then we consider the ESR and calculate the influence of 
anisotropy on the ESR linewidth in Sec.~\ref{sec:ESR}. 
Next the linear spin wave 
theory of the system is exploited under strong external fields in Sec.~\ref{sec:polarized} 
and the aspect of topological magnons is discussed. Finally in Sec.~\ref{sec:discussion} 
we comment on a possible material YbCl$_3$ and its potential realization of the Kitaev honeycomb model.

\section{Model}
\label{sec:model}

\begin{figure}[b]
	\centering
	\begin{tikzpicture}
	\draw (0:1) \foreach \x in {0,60,120,...,360} {-- (\x:1)};
	\draw[xshift=1.5cm,yshift=0.866cm] (0:1) \foreach \x in {60,120,...,360} {-- (\x:1)};
	\draw[xshift=1.5cm,yshift=-0.866cm] (0:1) \foreach \x in {60,120,...,360} {-- (\x:1)};
	\draw[thick,blue,->] (1,0)--(2,0);
	\node at (1.5,0.15) {$\mathbf{a}_3$};
	\draw[thick,green,->] (1,0)--(0.5,0.866);
	\node at (0.5,0.4) {$\mathbf{a}_1$};
	\draw[thick,red,->] (1,0)--(0.5,-0.866);
	\node at (1,-0.5) {$\mathbf{a}_2$};
	\draw[->] (-1.5,-1.5)--(-1.5,-0.5);
	\draw[->] (-1.5,-1.5)--(-0.5,-1.5);
	\draw (-1.5,-1.5) circle [radius=0.1];
	\filldraw[fill=black] (-1.5,-1.5) circle [radius=0.04];
	\node at (-0.7,-1.3) {$x$};
	\node at (-1.7,-0.7) {$y$};
	\node at (-1.8,-1.6) {$z$};
	\end{tikzpicture}
	\caption{The honeycomb lattice with three different types of bonds and our choice of the global coordinate system.}
	\label{fig:coordinates}
\end{figure}
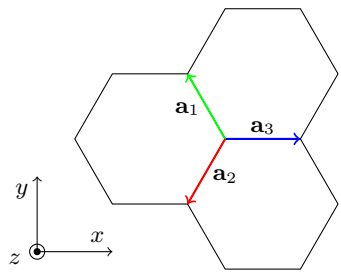

We begin with the following microscopic spin model, that is the most 
general nearest neighbor Hamiltonian on a honeycomb lattice with the 
(usual) {\sl Kramers} doublet effective spin-1/2 local 
moments~\cite{RauYb,YbMgGa2,PhysRevB.94.035107,PhysRevX.1.021002},
\begin{eqnarray}
\label{eq:Hamiltonian}
H &= & \sum_{\langle ij\rangle} J_{zz} S_i^z S_j^z + J_\pm (S_i^+S_j^- + S_i^-S_j^+)
\nonumber \\
& +& J_{\pm\pm} (\gamma_{ij} S_i^+ S_j^+ +\gamma_{ij}^* S_i^- S_j^-)
\nonumber \\
& +& J_{\pm z}[(\gamma_{ij}^* S_i^+ S_j^z+\gamma_{ij}S_i^-S_j^z)+\langle i \leftrightarrow j\rangle],
\end{eqnarray}
with $\gamma_{ij}$ taking $e^{2i\pi/3}$, $e^{-2i\pi/3}$, and $1$ on the bonds 
along $\mathbf{a}_1$, $\mathbf{a}_2$, $\mathbf{a}_3$ directions respectively, 
as shown in Fig.~\ref{fig:coordinates}. The spin components are defined in the 
global coordinate system in Fig.~\ref{fig:coordinates}. This is possible because 
the system is planar and has an unique rotational axis. This differs from the 
rare-earth pyrochlore materials where the spins are often defined in the 
local coordinate system for each sublattice. This model applies to the 
rare-earth local moment such as the Yb$^{3+}$ ion. 
For non-Kramers doublet like Pr$^{3+}$ or Tb$^{3+}$ ion, 
the $J_{\pm z}$ term is not allowed by symmetry, 
and the model becomes further simplified. 
In fact, a non-Kramers doublet based rare-earth honeycomb magnet arises 
from the triangular lattice magnet TbInO$_3$ after 1/3 of the Tb$^{3+}$ 
ions becomes inactive magnetically~\cite{Clark2019}. 
For the rare-earth local moments, the $4f$ electrons 
are much localized, and most often, one only needs to 
consider the nearest-neighbor interactions,
and occasionally, one would like to include the 
further neighbor dipole-dipole
interactions. In contrast, for the $4d/5d$ systems, 
one may need to worry about 
further neighbor exchange interactions because of 
the large spatial extension
of the electron wavefunctions.

An alternative and often used parametrization of the Hamiltonian 
is that of 
the $J$-$K$-$\Gamma$-$\Gamma'$ model~\cite{Rau}:
\begin{eqnarray}
\label{eq:Rau}
H&=& \sum_{\langle ij\rangle\in \alpha\beta(\gamma)}\left[ J \mathbf{S}_i\cdot \mathbf{S}_j+ K S_i^\gamma S_j^\gamma +\Gamma(S_i^\alpha S_j^\beta+S_i^\beta S_j^\alpha)\right]
\nonumber \\
& +& \Gamma'\sum_{\langle ij\rangle \in  \alpha\beta(\gamma) }\left(S_i^\alpha S_j^\gamma+S_i^\gamma S_j^\alpha + S_i^\beta S_j^\gamma + S_i^\gamma S_j^\beta\right),
 \end{eqnarray}
where $\alpha,\beta,\gamma$ take values in $\{x^\prime, y^\prime, z^\prime\}$. 
In the latter coordinate system, our unit vectors of Fig.~\ref{fig:coordinates} 
can be expressed by ${\hat{x}=(-1,-1,2)/\sqrt{6}}$, 
${\hat{y}=(1,-1,0)/\sqrt{2}}$ and ${\hat{z}=(1,1,1)/\sqrt{3}}$. 
The spin components in the above equation are 
\begin{equation}
\begin{split}
& S^{x^\prime}=-\frac{\sqrt{6}}{6} S_x +\frac{\sqrt{2}}{2} S_y + \frac{\sqrt{3}}{3} S_z,\\
& S^{y^\prime}=-\frac{\sqrt{6}}{6} S_x - \frac{\sqrt{2}}{2} S_y + \frac{\sqrt{3}}{3} S_z,\\
& S^{z^\prime}=\frac{\sqrt{6}}{3} S_x +\frac{\sqrt{3}}{3} S_z.\\
\end{split}
\end{equation}
Furthermore, we have used the notation that $\alpha\beta(\gamma)$ 
specifies a bond parallel (or anti-parallel) to the vector 
$\hat{{\alpha}}-\hat{{\beta}}$, or simply a bond of type $\gamma$. 
The coupling constants in Eq.~\eqref{eq:Hamiltonian} and \eqref{eq:Rau} 
are related by the following equation
\begin{equation}
\begin{split}
& J=\frac{4}{3}J_\pm-\frac{2\sqrt{2}}{3}J_{\pm z}-\frac{2}{3}J_{\pm\pm}+\frac{1}{3}J_{zz},\\
& K= 2\sqrt{2} J_{\pm z}+2 J_{\pm\pm},\\
& \Gamma=-\frac{2}{3}J_\pm-\frac{2\sqrt{2}}{3} J_{\pm z}+\frac{4}{3} J_{\pm\pm} +\frac{1}{3} J_{zz},\\
& \Gamma'=-\frac{2}{3}J_\pm+\frac{\sqrt{2}}{3}J_{\pm z}-\frac{2}{3}J_{\pm\pm}+\frac{1}{3}J_{zz}.
\end{split}
\end{equation}

The Hamiltonian in Eq.~\eqref{eq:Hamiltonian} can also be 
used to describe the general 
exchange interaction between the higher spin local moments 
for the honeycomb magnets after some 
modification.
The differences are explained in details in the Appendix~\ref{ssec1}.

\section{Thermodynamics}
\label{sec:thermo}

\begin{figure}[t]
	\centering
	\includegraphics[width=8.5cm]{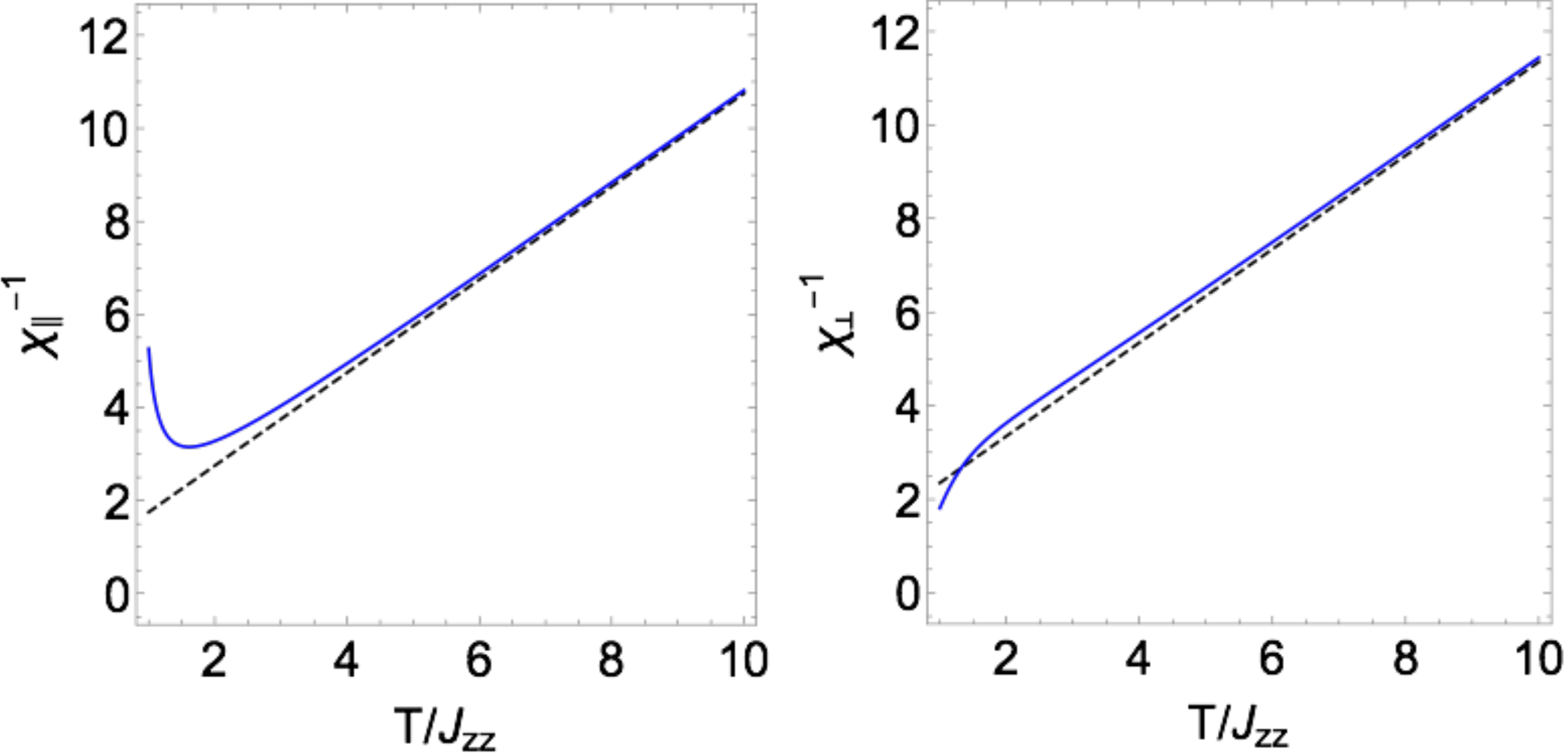}
	\caption{Susceptibilities versus temperature. The parameters are chosen to be $J_{zz}=1$, $J_\pm=0.9$, $J_{\pm\pm}=0.2,$ $J_{\pm z}=0.1$. The susceptibilities in the plot, $\chi_\parallel$ and $\chi_\perp$, are in units of ${\mu_0\mu_B^2 g_\parallel^2}/{4k_B}$ and ${\mu_0\mu_B^2 g_\perp^2}/{4k_B}$, respectively.}
	\label{fig:chi}
\end{figure}

The highly anisotropic nature of the exchange interaction first 
impacts the thermodynamic properties of the system. Here we explicitly
calculate the specific heat and the magnetic susceptibilities 
of the system from the generic exchange Hamiltonian. Using the 
high-temperature series expansion \cite{Fisher, HTEBook}, we find the heat capacity to be
\begin{equation}
C=\frac{3J_0^2}{2k_B T^2}-\frac{27 J_0^4}{8k_B^3 T^4},
\end{equation}
where we have 
\begin{equation}
J_0^2 \equiv \frac{1}{16} J_{zz}^2+\frac{1}{2} (J_\pm^2+J_{\pm\pm}^2+J_{\pm z}^2).
\end{equation}
Due to the spin-orbit entanglement, the coupling of the local moment 
to the external magnetic field is also anisotropic.
The Land\'e factors are different for the in-plane and out-plane
magnetic fields, and the Zeeman coupling is given as 
\begin{equation}
\label{eq:Zeeman}
H_Z=-\mu_0\mu_B \sum_i \left[ g_\perp (h_x S_i^x + h_y S_i^y)
+ g_\parallel h_\parallel S_i^z\right].
\end{equation}
Again using high-temperature series expansion, we compute the parallel 
and perpendicular spin susceptibilities up to $\mathcal{O}(T^{-3})$
\begin{equation}
\begin{split}
\chi_\parallel = \frac{\mu_0\mu_B^2 g_\parallel^2}{4k_B T} 
& 
\left(  1-\frac{3J_{zz}}{4k_B T}-\frac{ J_\pm^2}{2k_B^2T^2}-\frac{J_{\pm\pm}^2}{2k_B^2T^2}\right.\\
& \left.-\frac{J_{\pm z}^2}{k_B^2T^2}+\frac{3J_{zz}^2}{8k_B^2T^2}\right),\\
\chi_\perp = \frac{\mu_0\mu_B^2 g_\perp^2}{4k_B T} 
& 
\left( 1-\frac{3J_{\pm}}{2k_B T}+\frac{5 J_\pm^2}{4k_B^2T^2}-\frac{J_{\pm\pm}^2}{k_B^2T^2}\right.\\
& \left.-\frac{3J_{\pm z}^2}{4k_B^2T^2}-\frac{J_{zz}^2}{16k_B^2T^2}\right).
\end{split}
\end{equation}
In the SU(2)-symmetric point, ${J_{zz}=2J_{\pm}},$ 
${J_{\pm\pm}=J_{\pm z}=0}$, the two expressions coincide. For the rare-earth local moments with non-Kramers doublets, $g_\perp =0$ so 
$\chi_{\perp}=0$. 
In Fig.~\ref{fig:chi}, we plot the magnetic susceptibilities and 
show the deviation from the simple Curie-Weiss law due to the 
high order anisotropic terms. 

In addition to the simple thermodynamics such as $C_v$ and $\chi$, 
the magnetic torque measurement is proved to be quite useful in 
revealing the magnetic anisotropy. Intrinsically,
there is because the induced magnetization is generically 
not parallel to the magnetic field.
Thus, when the sample has an anisotropic magnetization, 
the system would experience 
a torque ${\tau=M\times H=-\partial F/\partial \theta}$ in an external magnetic field. 
The magnetotropic coefficient ${k=\partial^2 F/\partial \theta^2}$, defined as the second 
derivative of the free energy to the angle $\theta$ 
between the sample and the applied
magnetic field, can be introduced to quantify such anisotropy. 
It can be directly measured 
using the resonant torsion magnetometry~\cite{Torque}. Under the high temperature expansion, 
we find the magnetotropic coefficient $k$ is given as 
\begin{equation}
\begin{split}
k=& \frac{\mu_0^2 \mu_B^2 h^2}{k_B T}\cos2\theta\big\{ \frac{1}{4}(g_\perp^2-g_\parallel^2)+\frac{3}{16k_BT}(g_\parallel^2J_{zz}-2g_\perp^2J_\pm)\\
& +\frac{1}{192k_B^2T^2}\big[-3g_\perp^2(-20J_{\pm}^2+16J_{\pm\pm}^2+12J_{\pm z}^2+J_{zz}^2)\\
& +6g_\parallel^2(4J_\pm^2+4J_{\pm\pm}^2+8J_{\pm z}^2-3J_{zz}^2)\\
& -2\mu_0^2\mu_B^2h^2(g_\perp^4-g_\parallel^4)\big]\big\}
-\frac{\mu_0^4\mu_B^4h^4}{96k_B^3T^3}\cos4\theta(g_\perp^2-g_\parallel^2)^2,
\end{split}
\end{equation}
where we have defined $h^2=h_x^2+h_y^2+h_z^2.$ The coefficient $k$ vanishes in the g-isotropic $g_\perp=g_\parallel$ and Heisenberg limit: ${J_{zz}=2J_{\pm}}$, 
${J_{\pm\pm}=J_{\pm z}=0}$. More details of the calculation can be found 
in Appendix~\ref{ssec2}. 

\begin{figure*}
	\centering
\includegraphics[width=5.8cm]{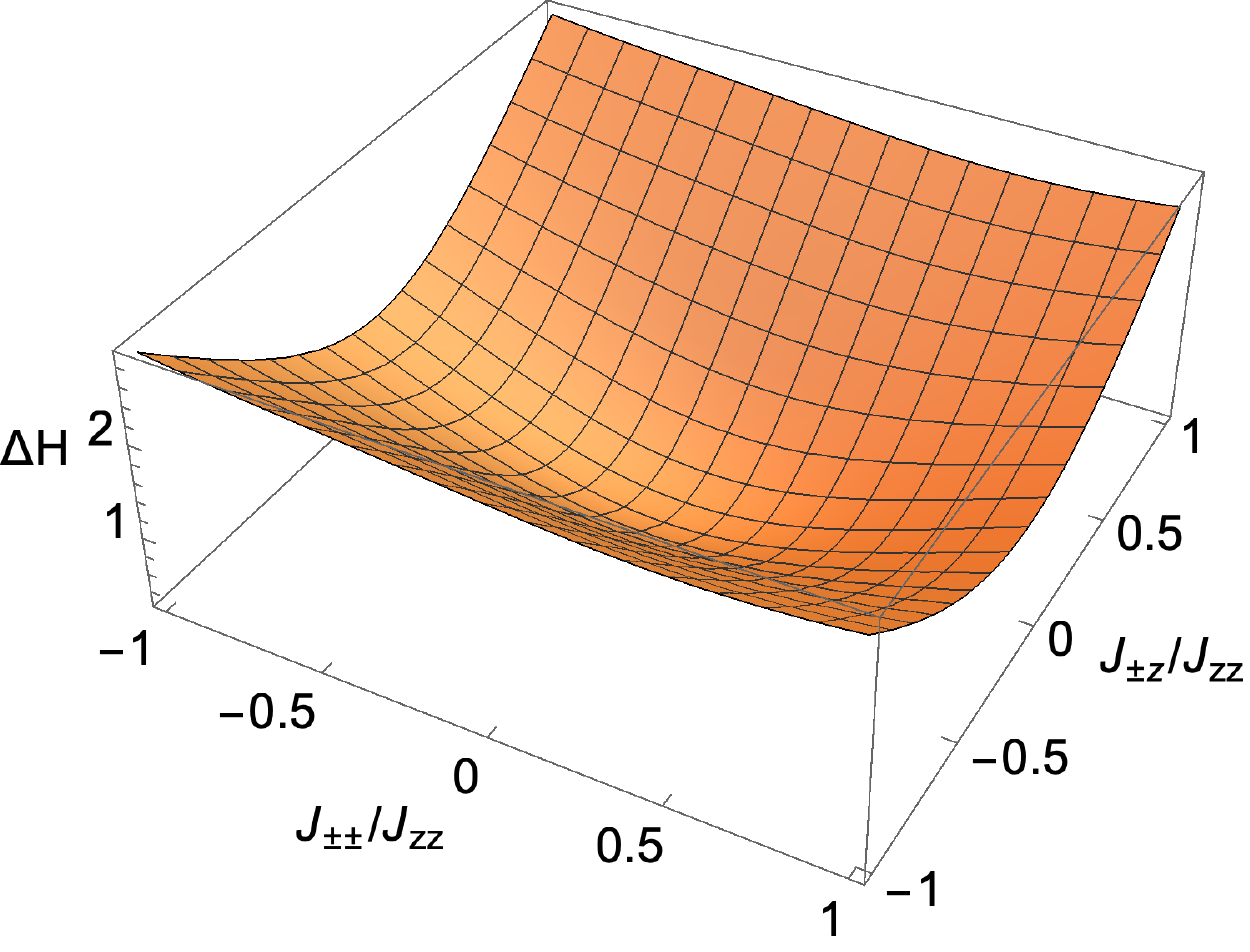}
\includegraphics[width=5.8cm]{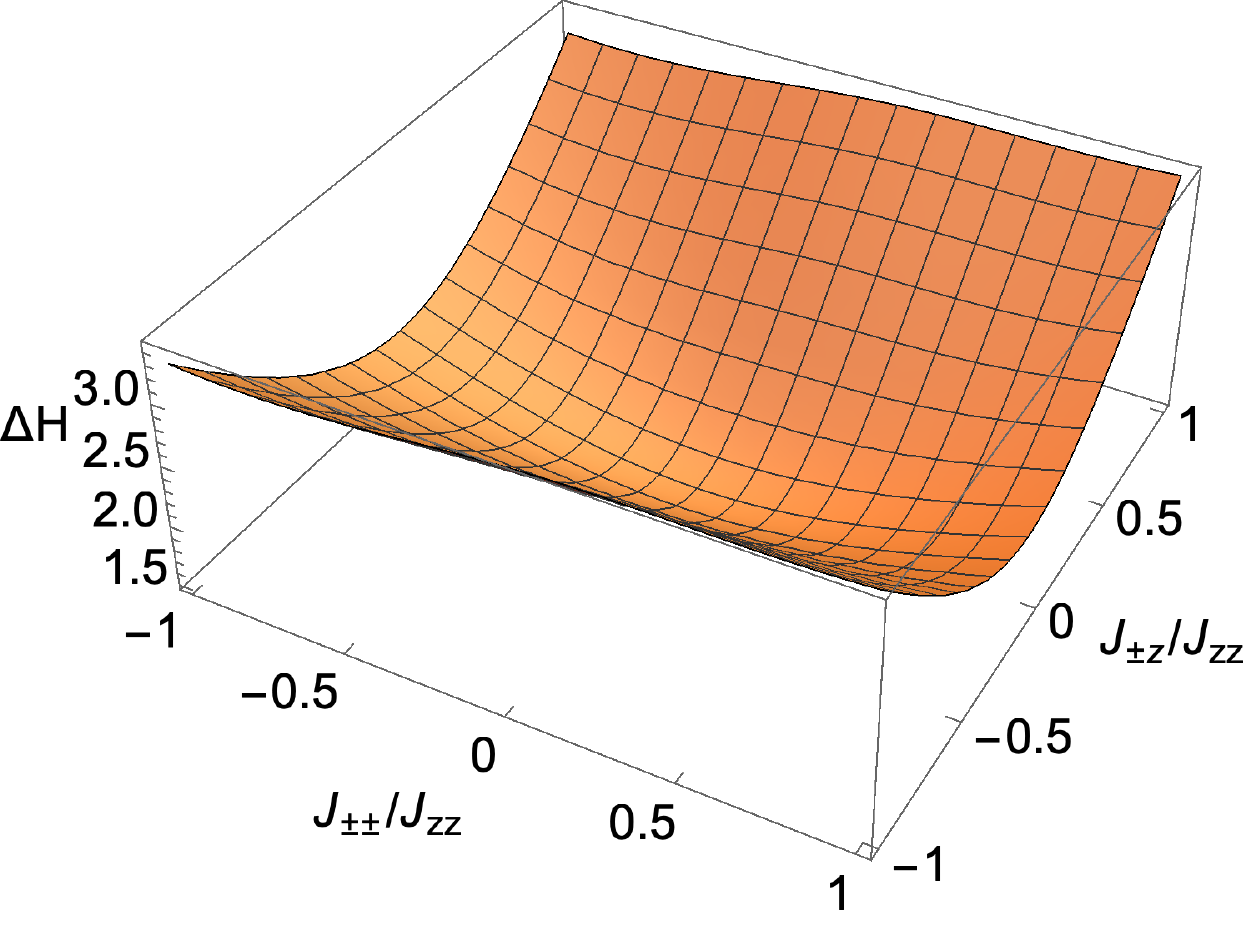}
\includegraphics[width=5.8cm]{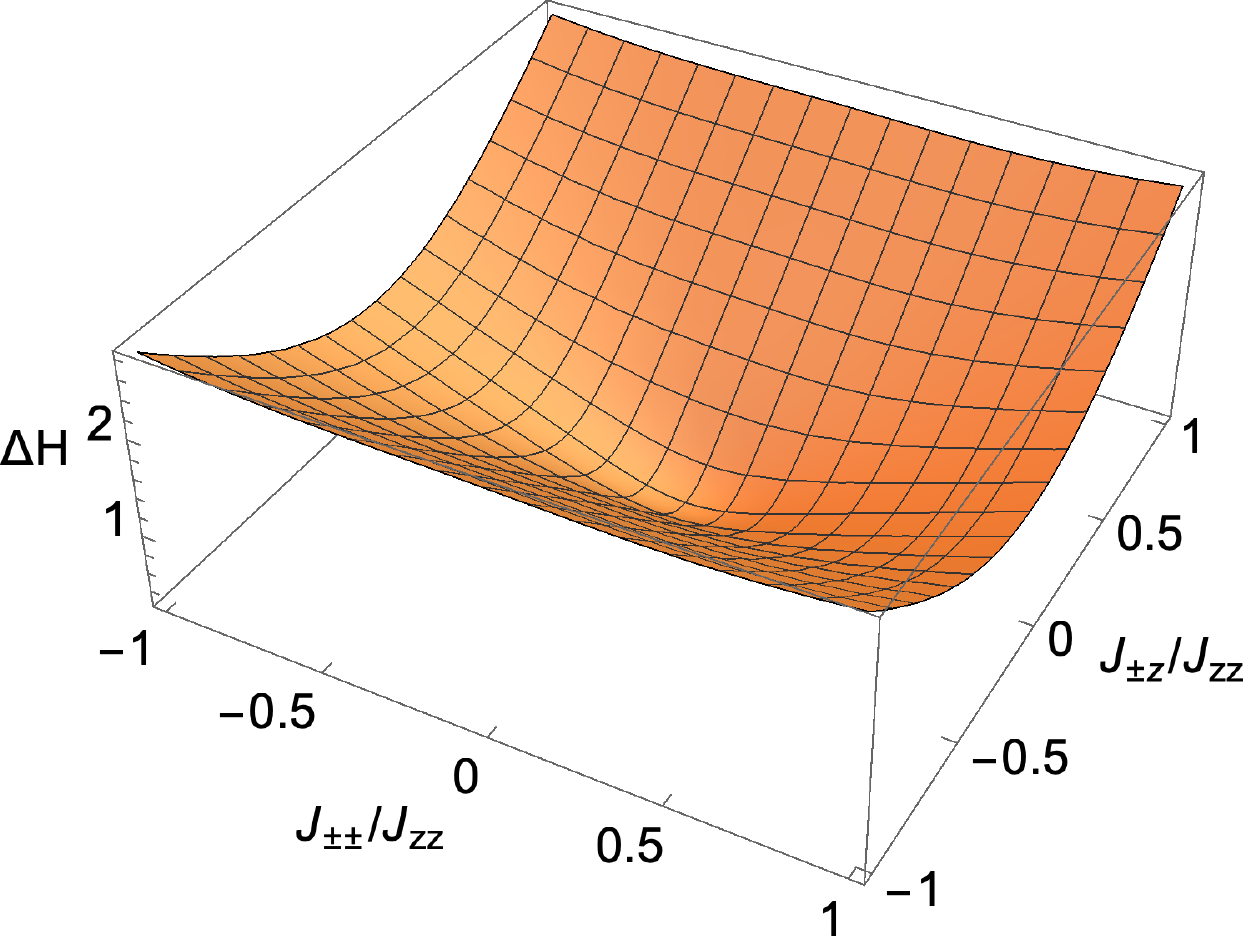}
\caption{(Color online.) The dependence of $\Delta H$ on $J_{\pm\pm}$ and $J_{\pm z}$. 
Left:  $J_{\pm}/J_{zz}=1$, middle: $J_{\pm}/J_{zz}=-0.5$, right: $J_{\pm}/J_{zz}=0.2$. 
The linewidth $\Delta H$ in the three plots are in the units $\mu_B g(\theta) /\sqrt{2\pi}$.}
\label{fig:ESR}
\end{figure*}

\section{Electron spin resonance}
\label{sec:ESR}

In the thermodynamic properties, the leading contributions 
come from the $J_{zz}$ and $J_\pm$ terms, while the $J_{\pm\pm}$ 
and $J_{\pm z}$ terms are subleading. Arising from spin-orbital 
entanglement and completely breaking the U(1) rotational symmetry, 
these terms play important roles in the potential quantum spin liquid 
behavior. To resolve them, we now turn to the electron spin resonance.

Electron spin resonance measures the absorption of electromagnetic radiation 
by a sample subjected to an external static magnetic field. For a SU(2) 
invariant system, the absorption is completely sharp, {\sl i.e.} described by 
a delta function located exactly at the Zeeman energy~\cite{Oshikawa}. 
Therefore, the broadening of the resonance spectrum has to arise from 
the magnetic anisotropy. To understand the contribution of the anisotropy 
of the nearest-neighbor spin interaction to the ESR linewidth, 
we decompose the Hamiltonian Eq.~\eqref{eq:Hamiltonian} into 
the isotropic Heisenberg part and the anisotropic exchange part
\begin{equation}
H=J \sum_{\langle i,j\rangle} \mathbf{S}_i \cdot \mathbf{S}_j + H',
\end{equation}
where the Heisenberg coupling $J=(J_{zz}+4J_{\pm})/3$, and the anisotropic part 
\begin{equation}
H'=\sum_{\langle i, j\rangle } S_i^\mu \, \Gamma_{ij,\mu\nu} \, S_j^\nu.
\end{equation}
Here $\Gamma_{ij}$ a traceless and symmetric exchange coupling matrix, satisfying 
\begin{equation}
\begin{split}
& \Gamma_{ij,xx}=2J_\pm/3+(\gamma_{ij}+\gamma_{ij}^*)J_{\pm\pm}-J_{zz}/3,\\ 
& \Gamma_{ij,yy}=2J_\pm/3-(\gamma_{ij}+\gamma_{ij}^*)J_{\pm\pm}-J_{zz}/3, \\
& \Gamma_{ij,xy}= i(\gamma_{ij}-\gamma_{ij}^*)J_{\pm\pm},\\
& \Gamma_{ij,yz}=i(\gamma_{ij}^*-\gamma_{ij})J_{\pm z},\\
& \Gamma_{ij,zx}=(\gamma_{ij}+\gamma_{ij}^*)J_{\pm z}.
\end{split}
\end{equation}

Under the Zeeman term of Eq.~\eqref{eq:Zeeman}, the ESR linewidth 
for a Lorentzian-shaped spectrum is~\cite{Linewidth, Castner, Soos}
\begin{equation}
\Delta H(\theta)=\frac{\sqrt{2\pi}}{\mu_B g(\theta)}\left(\frac{M_2^3}{M_4}\right)^{1/2} ,
\end{equation}
where $\theta$ is again the angle between the external field 
and the sample, 
and 
\begin{eqnarray}
g(\theta)&=&\sqrt{g_\parallel^2\sin^2\theta+g_\perp^2\cos^2\theta} , \\
M_2 &=& \frac{ \langle [H', M^+][M^-,H']\rangle} {\langle M^+ M^-\rangle} , \\
M_4 &=& \frac {\langle [H,[H',M^+]][H,[H',M^-]]\rangle } {\langle M^+ M^-\rangle}.
\end{eqnarray}
$M_2$ and $M_4$
are the second and the fourth moments, respectively, and 
$M^\pm\equiv \sum_i S_i^\pm$. 
The expectation ``$\langle \cdots \rangle$''in the above equations 
is taken with respect to high temperatures.
Specifically, we find that 
\begin{equation}
\begin{split}
M_2=& \frac{3}{4}(J_{zz}^2+4J_\pm^2+4J_{\pm\pm}^2+10J_{\pm z}^2-4J_{\pm}J_{zz}),\\
M_4= & \frac{3}{4}J_{zz}^4-\frac{9}{2}J_{zz}^3J_\pm
+\frac{57}{8}J_{zz}^2J_{\pm z}^2
+15 J_{zz}^2 J_{\pm}^2\\
&+6J_{zz}^2J_{\pm\pm}^2
-\frac{3}{4}J_{zz}J_{\pm\pm}J_{\pm z}^2
-\frac{93}{4}J_{zz}J_{\pm} J_{\pm z}^2\\
&-24 J_{zz} J_{\pm} J_{\pm\pm}^2
-30J_{zz}J_{\pm}^3
+\frac{123}{2}J_{\pm z}^4\\
& +\frac{153}{2}J_{\pm\pm}^2J_{\pm z}^2
+\frac{39}{2}J_{\pm}J_{\pm\pm}J_{\pm z}^2
+33J_{\pm}^2J_{\pm z}^2\\
& +15 J_{\pm\pm}^4
+30J_\pm^2 J_{\pm\pm}^2
+24J_\pm^4.
\end{split}
\end{equation}

Our result for ESR linewidths can be compared to the future ESR experiments
on the rare-earth based honeycomb magnets in order to extract the anisotropic exchanges. In Fig.~\ref{fig:ESR}, we further
depict the three-dimensional plots that explicitly demonstrate 
the dependence of the ESR linewidth on the anisotropic couplings 
$J_{z\pm}$ and ${J_{\pm\pm}}$ for three different choices of $J_{\pm}$.

\section{Polarized phases}
\label{sec:polarized}

\subsection{Strong field normal to the honeycomb plane}

To further explore the effect of the anisotropic exchange interaction, we study the spin wave excitation with respect to the polarized states under the strong magnetic fields. This is clearly feasible in the current laboratory setting for the rare-earth magnets as the energy scales for them are usually rather small. For the $4d/5d$ magnets, there can be difficulty to achieve as the energy scale over there is much higher. Our results here are relevant to the inelastic neutron scattering and thermal Hall transport measurements.

We first consider the case of a strong magnetic field in the direction normal
to the honeycomb plane such that the system is in the fully polarized paramagnetic 
phase and all the spins are aligned along the $z$ direction. In this case, the magnon bands carry nontrivial Chern numbers for generic range of parameters, as found in reference \cite{McClarty} in the $J-K-\Gamma-\Gamma'$ presentation.   
We expand about this fully polarized state using the conventional Holstein-Primakoff 
transformations of the spin variables~\cite{Holstein}, 
which are ${S_i^z=S-a_i^\dagger a_i, 
 S_i^+= a_i, 
 S_i^-= a_i^\dagger} $
for sublattice A, and substitute $a\rightarrow b$ for sublattice B. 
$a$ and $b$'s are bosonic operators, $[a_i, a_j^\dagger]=[b_i,b_j^\dagger] 
=\delta_{ij}$. Keeping only the 
bilinear terms of bosonic operators and taking the Fourier transformation, we arrive at
\begin{equation}
H= \frac{3N}{4} J_{zz}-2N \mu_0\mu_B g_\parallel h_z+\frac{1}{2} 
\Upsilon_\k^\dagger \mathcal{H}_\k \Upsilon_\k,\
\end{equation}
with ${\Upsilon \equiv (a_\k,b_\k,a_{-\k}^\dagger,b_{-\k}^\dagger)^T}$.
Here we have
denoted ${k_1=-\frac{1}{2}k_x+\frac{\sqrt{3}}{2}k_y},{k_2=-\frac{1}{2}k_x-\frac{\sqrt{3}}{2}k_y}$, 
and ${k_3=k_x}$
that correspond to the $y$-, $z$- and $x$-bonds, respectively. We further 
define ${f(\k)= \sum_i e^{i k_i}}$, ${g_1(\k)=\sum_i e^{-i k_i} \gamma_i}$, 
${g_2(\k)=\sum_i e^{i k_i} \gamma_i}$ and $u\equiv( g_\parallel\mu_0\mu_B h_\parallel-3J_{zz}/2)$,
we then have for $\mathcal{H}_\k$ a block form
\begin{equation}
\mathcal{H}_{\mathbf{k}}=
\left[\begin{matrix} A(\mathbf{k}) & B(\mathbf{k}) \\ 
B^\dagger(\mathbf{k}) & A^T(-\mathbf{k})\end{matrix}\right],
\end{equation}
where we have 
\begin{eqnarray}
A(\mathbf{k})&=& \left[\begin{matrix} u & J_\pm f^* \\ J_\pm f & u\end{matrix}\right], \\
B^\dagger(\mathbf{k})&=&\left[ \begin{matrix} 0 & J_{\pm\pm}g_1 \\ J_{\pm\pm} g_2 & 0\end{matrix}\right].
\end{eqnarray}
All the $J_{\pm z}$ terms are not present. 

The spin wave dispersion relation for $\mathcal{H}_\k$ follows as
\begin{equation}
\begin{split}
\epsilon(\k)^2=&  u^2+ |f|^2 J_\pm^2-\frac{|g_1|^2+|g_2|^2}{\sqrt{2}}J_{\pm\pm}^2\\
&\pm [4|f|^2u^2J_\pm^2 +\frac{|g_1|^2-|g_2|^2}{4} J_{\pm\pm}^4 \\
& +(f^*g_1^*-fg_2^*)(f^*g_2-fg_1)J_\pm^2J_{\pm\pm}^2 ]^{1/2},
\end{split}
\end{equation}
where only the positive square root of $\epsilon (\k)^2$ is taken.  

Several simple limits of this expression can be checked: (1) in the Heisenberg limit $J_{zz}=2J_{\pm}\equiv 2J$, then $\epsilon(\mathbf{k})=(\frac{g_\parallel \mu_0\mu_B}{2S} h_\parallel-3J \pm |f| J)^{1/2}$; (2) when only $J_{zz}$ is finite, it reduces to the Ising case $\epsilon(\mathbf{k})=\frac{g_\parallel \mu_0\mu_B}{2S} h_\parallel-\frac{3}{2}J_{zz}$; if only $J_{\pm}$ is present, we have a graphene-like dispersion $\epsilon(\mathbf{k})=\frac{g_\parallel \mu_0\mu_B}{2S} h_\parallel \pm |f| J_\pm$.

\begin{figure}[t]
	\centering
	\includegraphics[width=8cm]{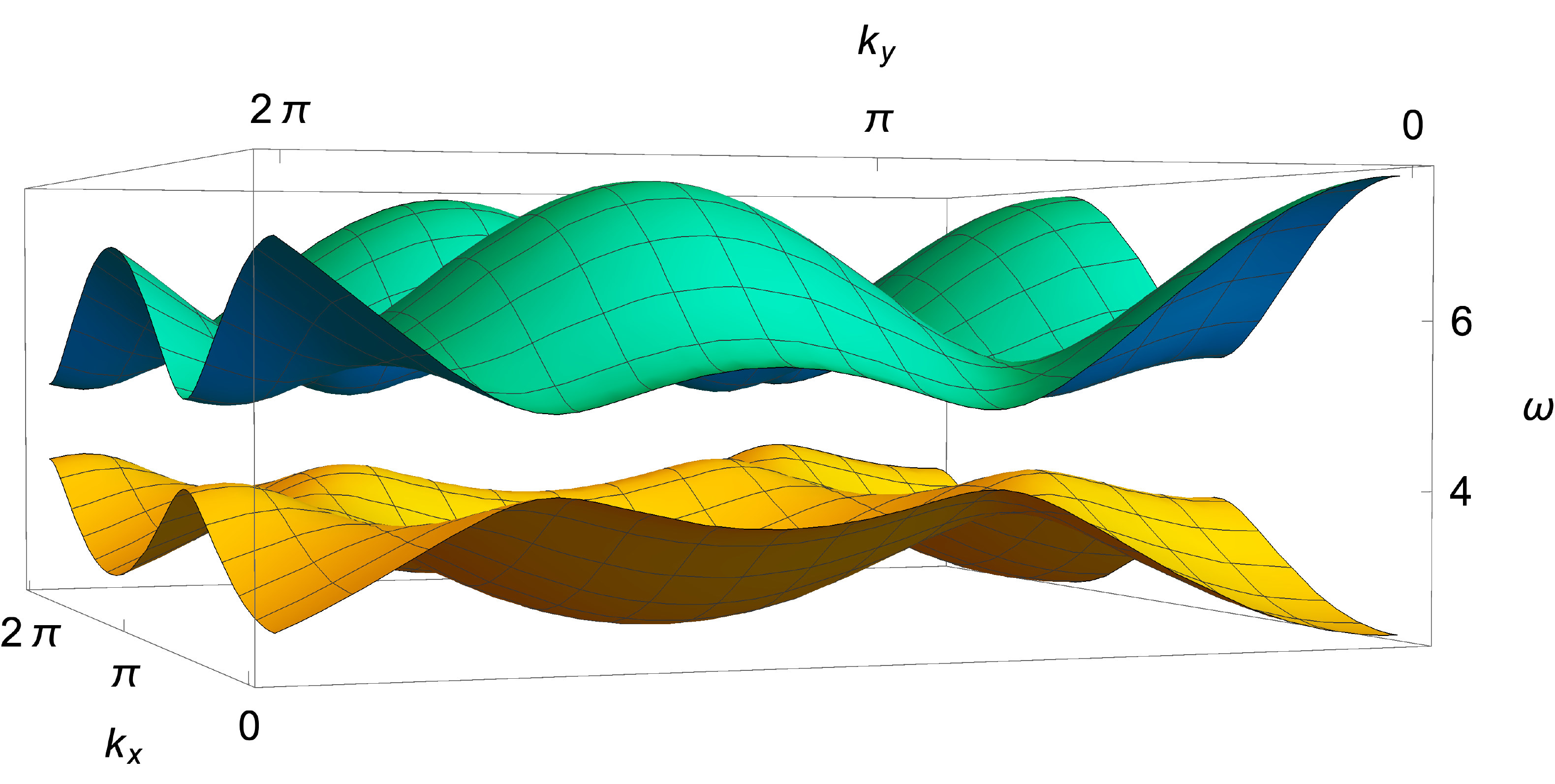}
	\caption{(Color online.) The two spin wave bands $\omega_\pm$ 
	when a strong field in the $z$-direction is applied. 
	The parameters are chosen as ${J_{zz} = 1; J_\pm = 0.9; J_{\pm\pm} = 1; J_{\pm z} = 0.3; u = 5}$.}
	\label{fig:SpinWaveZ}
\end{figure}

At high fields, the results can be simplified by the Schrieffer-Wolff transformation,
\begin{eqnarray}
\tilde{\mathcal{H}}_{\mathbf{k}}&=&e^W \mathcal{H}_{\mathbf{k}} e^{-W}
\nonumber \\
&=&\mathcal{H}_{\mathbf{k}}+[W,\mathcal{H}_{\mathbf{k}}]+\frac{1}{2}\big[W,[W,\mathcal{H}_{\mathbf{k}}]\big]+\cdots,
\end{eqnarray}
with the commutator understood as 
\begin{equation}
[X,Y]\equiv X \eta Y- Y\eta X ,
\end{equation}
and $\eta$ is a diagonal matrix with entries $(1,1,-1,-1)$. 
Following the treatment of Ref.~\cite{McClarty}, we 
choose the transformation to be 
\begin{equation}
W=\frac{1}{2u}\left(\begin{matrix} 0 & B(\mathbf{k})\\ 
-B^\dagger(\mathbf{k}) & 0\end{matrix}\right),
\end{equation}
so that up to $O(h_\parallel^{-2})$, we have the 
$\tilde{\mathcal{H}}_{\mathbf{k}}$ 
to become $A(\mathbf{k})\rightarrow \tilde{A}(\mathbf{k})$, 
$B(\mathbf{k})\rightarrow \tilde{B}(\mathbf{k})$,
\begin{equation}
\begin{split}
& \tilde{A}(\mathbf{k})
=\left(\begin{matrix} u-\frac{J_{\pm\pm}^2}{2u}|g_2|^2 & f^* J_\pm \\ 
f J_\pm & u-\frac{J_{\pm\pm}^2}{2u}|g_1|^2\end{matrix}\right),\\
& \tilde{B}(\mathbf{k})
= -\frac{J_\pm J_{\pm\pm}}{2u}\left(\begin{matrix}f^*g_1^*+g_2^*f & 0 \\ 
0 & f^*g_1^*+g_2^*f\end{matrix}\right).
\end{split}
\end{equation}
At high fields, we can thus ignore $\tilde{B}(\mathbf{k})$ and 
focus on the $\tilde{A}(\mathbf{k})$ term. Writing  
$\tilde{A}(\mathbf{k})=d_0(\mathbf{k})\mathbbm{1}
+\frac{1}{2}\mathbf{d}(\mathbf{k})\cdot {\boldsymbol{\sigma}},$
with the three components being
\begin{eqnarray}
 d_1(\mathbf{k}) &= & 2J_\pm~Re(f),\\
 d_2(\mathbf{k}) &=& 2J_\pm~Im(f), \\
 d_3(\mathbf{k}) &=&\frac{J_{\pm\pm}^2}{2u}(|g_1|^2-|g_2|^2), \\
 d_0(\mathbf{k}) &=&u-\frac{J_{\pm\pm}^2}{4u}(|g_1|^2+|g_2|^2). 
\end{eqnarray}

\begin{figure}[t]
\centering
\includegraphics[width=6.5cm]{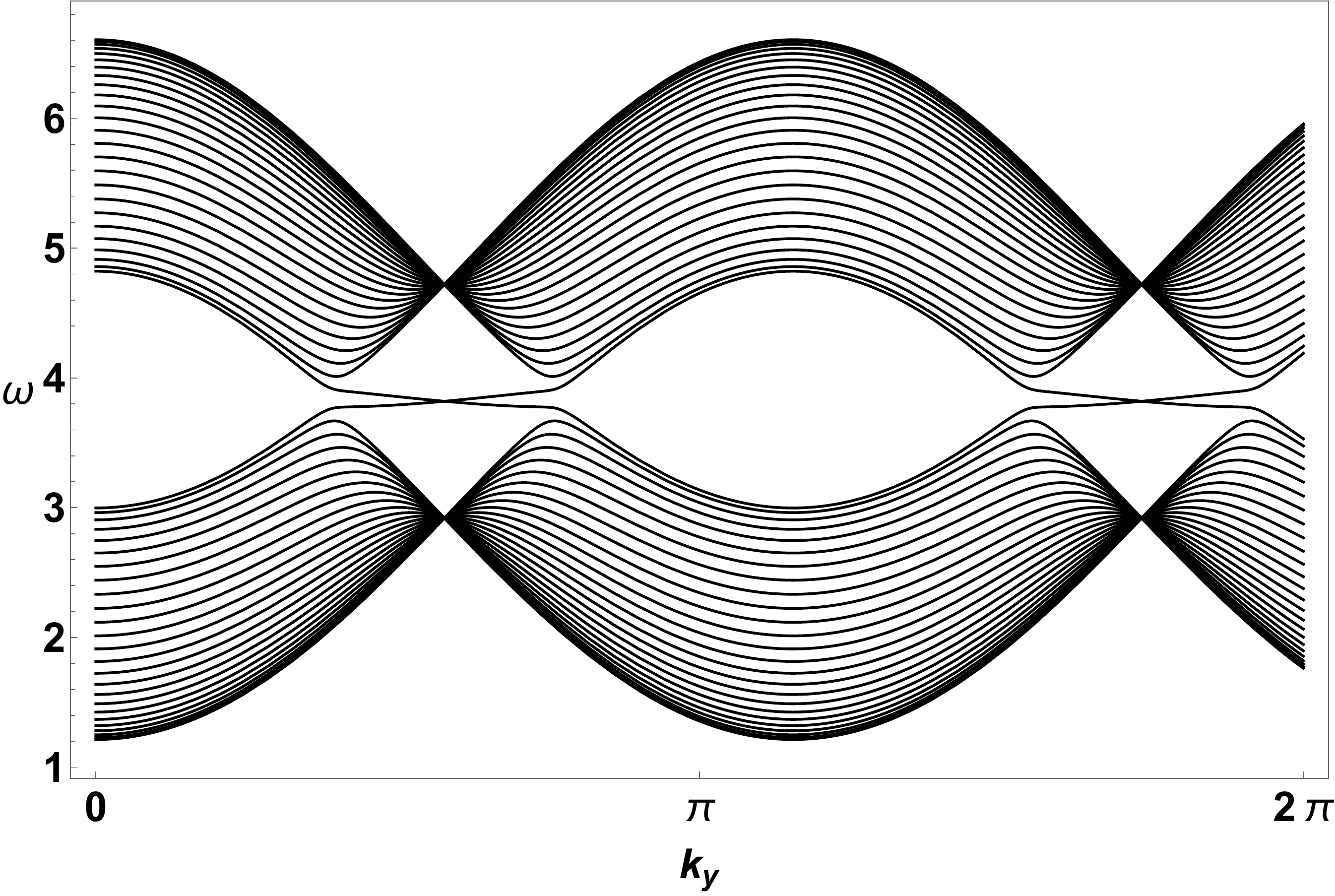}
\caption{Edge state in a cylindrical geometry. The $y$-direction is periodic, 
while $x$-direction contains 50 sites. The parameters are 
$J_{zz}=1,$ $J_{\pm} = 0.9,$ $J_{\pm\pm} = 0.6,$ $u = 4.$}
	\label{fig:edge}
\end{figure}

At each momentum $\mathbf{k}$ we have the eigenvalues
\begin{equation}
\omega_\pm(\mathbf{k})=d_0(\mathbf{k})\pm \frac{1}{2}|\mathbf{d}(\mathbf{k})|.
\label{eq:dispersion}
\end{equation}
The above spin wave bands Eq.~\eqref{eq:dispersion} do not touch 
unless $J_\pm=J_{\pm\pm}=0$, as we have depicted in
Fig.~\ref{fig:SpinWaveZ}.
We further compute the Berry curvature as follows
\begin{equation}
\label{eq:Berry}
F_\pm^{xy}(\mathbf{k})=\pm\frac{i}{2} \left[\frac{\mathbf{d}(\mathbf{k})}{|\mathbf{d}(\mathbf{k})|^3}\cdot\left(\frac{\partial \mathbf{d}(\mathbf{k})}{\partial k_y}\times\frac{\partial \mathbf{d}(\mathbf{k})}{\partial k_x}\right)\right].
\end{equation}
This is negative semi-definite in the Brillouin zone. 
The Chern numbers follow as
\begin{equation}
C_\pm=\frac{1}{2\pi i}\int_{BZ} dk_x dk_y F_{\pm}^{xy}=\mp 1.
\end{equation}
This implies the presence of chiral magnon edge states and thermal Hall effect, resulting from the presence of magnon number non-conserving terms $B(\k)$ in the Hamiltonian~\cite{Matsumoto, McClarty}. 
The edge state for the open boundary condition is depicted in Fig.~\ref{fig:edge}.

\subsection{Strong field in the honeycomb plane}

We now turn to a strong in-plane field, in the $x$-direction. This is relevant for 
the rare-earth local moments with the usual Kramers doublet, and does not apply to the
non-Kramers doublet. 
The Holstein-Primakoff transformation for sublattice A is modified as $
S_i^x=\frac{1}{2}-a_i^\dagger a_i, S_i^y=\frac{1}{2}(a_i+a_i^\dagger),  
S_i^z=\frac{1}{2i}(a_i-a_i^\dagger)$, and that for sublattice B is 
obtained by substituting $a$ by $b$. These are again bosonic operators 
satisfying $[a_i,a_j^\dagger]=[b_i,b_j^\dagger]=\delta_{ij}$.
Keeping only the 
bilinear terms of bosonic operators and taking the Fourier transformation, 
we obtain
\begin{eqnarray}
 H &=& \frac{3N}{2} J_{\pm}-2N \mu_0\mu_B g_\perp h_x
        + \frac{1}{2} \Upsilon_\k^\dagger \mathcal{H}_\k \Upsilon_\k,
   \\
\Upsilon&\equiv& (a_\k,b_\k,a_{-\k}^\dagger,b_{-\k}^\dagger)^T.
 \end{eqnarray}
Define $g_3(\mathbf{k})=e^{ik_1}+e^{ik_2}-2e^{ik_3}$, 
$g_4(\mathbf{k})=e^{ik_1}-e^{ik_2}$ 
and recall $f(\k)= \sum_i e^{i k_i}$. 
The $\mathcal{H}_k$ is of the familiar form
\begin{equation}
\mathcal{H}_{\mathbf{k}}=\left[\begin{matrix} A(\mathbf{k}) & B(\mathbf{k}) \\ 
B^\dagger(\mathbf{k}) & A^T(-\mathbf{k})\end{matrix} \right],\nonumber
\end{equation}
but now with the $A, B$ matrices given by 
\begin{eqnarray}
&& A(\mathbf{k})_{11}=A(\mathbf{k})_{22}=v=\mu_0\mu_Bg_\perp h_x-3J_\pm , \\
&& A(\mathbf{k})_{21}=A(\mathbf{-k})_{12}=(\frac{1}{4}J_{zz}+\frac{1}{2}J_{\pm})f+\frac{g_3}{4}J_{\pm\pm}, \\
&& B(\mathbf{k})_{11}=B(\mathbf{k})_{22}=0, \\
&& B(\mathbf{k})_{21}=B(\mathbf{-k})_{12}=(-\frac{1}{4}J_{zz}+\frac{1}{2}J_\pm)f
\nonumber \\
&& \quad \quad \quad \quad \quad \quad \quad +\frac{1}{4}J_{\pm\pm}g_3+\frac{i\sqrt{3}}{2}J_{\pm z}g_4 . 
\end{eqnarray}

\begin{figure}[t]
\centering
\includegraphics[width=8cm]{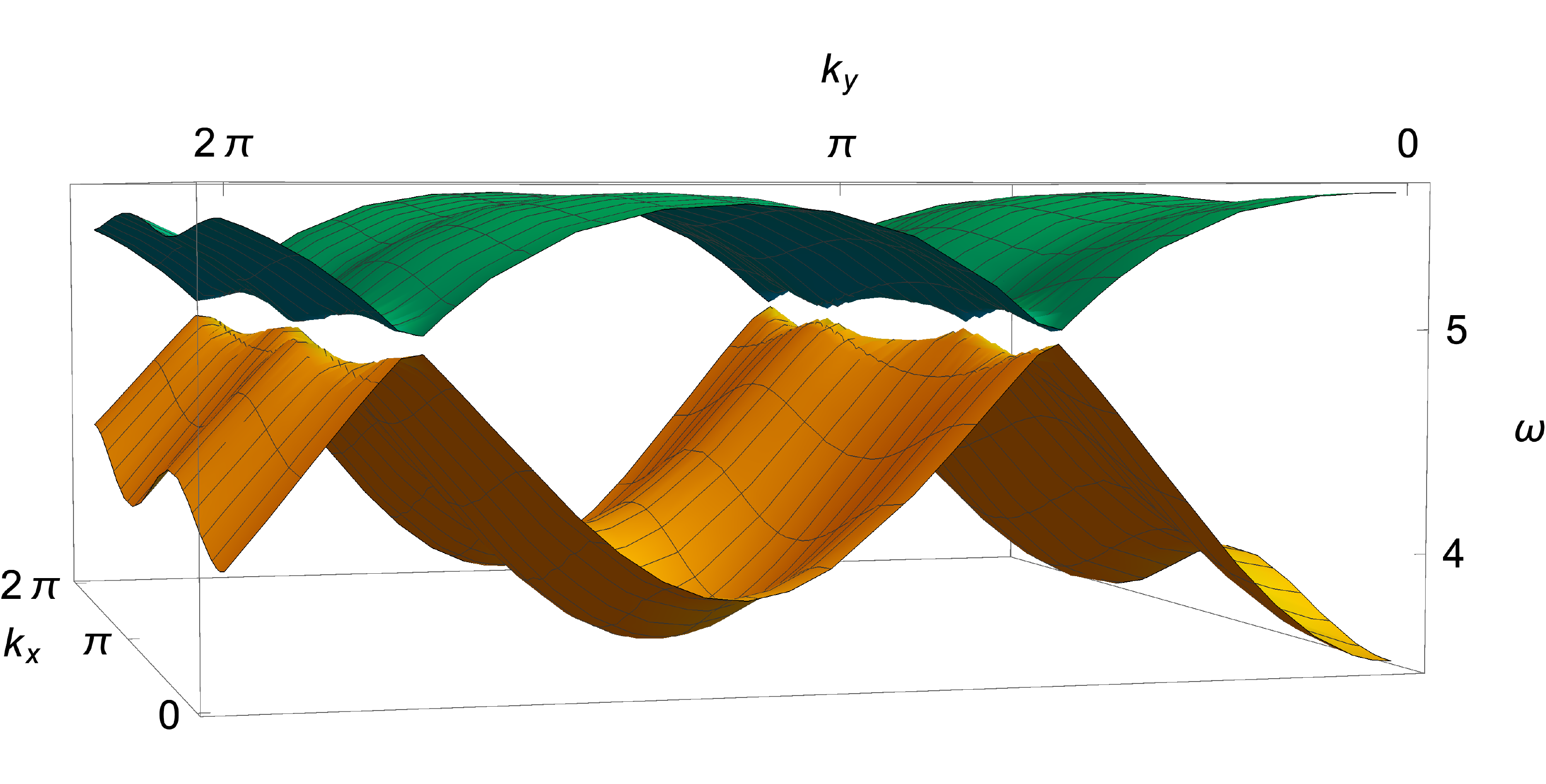}
\caption{(Color online.) The two spin wave bands $\omega_\pm$ when a strong in-plane field is present. 
The parameters are chosen $J_{zz} = 1$; $J_\pm = 0.9$; $J_{\pm\pm} = 1$; $J_{\pm z } = 0.3$; $ v = 5$.}
\label{fig:SpinWaveX}
\end{figure}

Appealing again to the Schrieffer-Wolff transformation with
\begin{equation}
W=\frac{1}{2v}\left(\begin{matrix} 0 & B(\mathbf{k})\\ -B^\dagger(\mathbf{k}) & 0\end{matrix}\right),
\end{equation}
then up to $O(h_\perp^{-2})$, we have the effective $\tilde{\mathcal{H}}_k$ to be $A(\mathbf{k})\rightarrow \tilde{A}(\mathbf{k})$, $B(\mathbf{k})\rightarrow \tilde{B}(\mathbf{k})$.
\begin{eqnarray}
&& \tilde{A}(\mathbf{k})=\left(\begin{matrix} v-\frac{1}{2v}|B(\mathbf{k})_{12}|^2 & A(\mathbf{k})_{12}\\ A(\mathbf{k})_{21} & v-\frac{1}{2v}|B(\mathbf{k})_{21}|^2\end{matrix}\right),\\
&& \tilde{B}(\mathbf{k})=-\frac{ \mathbbm{1} }{2v} \big[ A(\mathbf{k})_{21}B(\mathbf{k})_{12}+A(\mathbf{k})_{12}B(\mathbf{k})_{21 } \big].
 \end{eqnarray}
At high fields $h_\perp$, we can ignore $\tilde{B}(\mathbf{k})$ and focus on the $\tilde{A}(\mathbf{k})$ term. Rewrite $\tilde{A}(\mathbf{k})=d_0(k)\mathbbm{1}+\frac{1}{2}\mathbf{d}(\mathbf{k})\cdot \mathbf{\sigma}$, 
with each component being
\begin{eqnarray}
&& d_1=(\frac{1}{4}J_{zz}+\frac{1}{2} J_\pm)Re(f)+\frac{1}{4}J_{\pm\pm}Re(g_3),\\
&& d_2=(\frac{1}{4}J_{zz}+\frac{1}{2} J_\pm)Im(f)+\frac{1}{4}J_{\pm\pm}Im(g_3),\\
&& d_3=-\frac{1}{2v}\big[i\frac{\sqrt{3}}{2}g_4^* J_{\pm z} [(\frac{1}{4}J_{zz}+\frac{1}{2} J_\pm)f+\frac{1}{4}J_{\pm\pm}g_3] \nonumber \\
&&\quad\quad\quad\quad\quad\quad \quad\quad\quad+ c.c\big], \\
&& d_0=v-\frac{1}{2v}[(\frac{1}{4}J_{zz}+\frac{1}{2} J_\pm)^2|f|^2+\frac{3}{4}J_{\pm z}^2|g_4|^2].
\end{eqnarray} 
We then arrive at the dispersions 
$\omega_\pm(\mathbf{k})=d_0(\mathbf{k})\pm \frac{1}{2} |\mathbf{d}(\mathbf{k})|$. 
The spectrum is plotted in Fig.~\ref{fig:SpinWaveX}.  
We find both bands have zero Chern numbers, and we have checked for many
other parameter choices and also obtained trivial zero Chern number. Thus,
the in-plane field magnon band structure is quite distinct from the 
topological magnon band structure for the normal-plane field case.

\section{Discussion}
\label{sec:discussion}

%The rare-earth based honeycomb Kitaev-like materials have not been carefully explored yet. These systems can be equally interesting as the $4d/5d$ magnets with the ${J_{\text{eff}}=1/2}$ local moments, and the anisotropic spin interaction can be more significant than the $4d/5d$ magnets due to the much stronger spin-orbit couplings in these materials. In our work here, 

We have studied the experimental consequences of the spin-orbital entanglement and the anisotropic spin exchange interactions in the honeycomb rare-earth magnets. These results can be directly compared with the experiments, thereby providing a useful guidance for the future study on candidate systems. One future direction would be to involve higher-lying crystal field states based on the information of specific materials. 

One potential rare-earth candidate for the anisotropic honeycomb lattice model is YbCl$_3$ \cite{Chemistry}, which has a similar crystal structure to that of RuCl$_3$. The Yb$^{3+}$ ions have nearly filled $4f$-orbitals, which, combined with the large crystal fields lead to Kramers doublet ground state manifold. This is modeled as an effective spin-1/2 local moment. Furthermore, its edge-shared octahedral structure gives simple exchange physics that is relatively well-understood according to a microscopic calculation in Ref.~\onlinecite{RauYb}. There is very limited information about this material in the literature apart from a very recent work \cite{Ni}. 

In this paper, we have focused the analysis on the honeycomb lattice rare-earth magnets and its anisotropic interaction. It is noticed that, the generic model for the rare-earth honeycomb magnets contains a Kitaev interaction as one independent exchange interaction out of four. It is thus reasonable for us to consider the possibility of Kitaev materials among the honeycomb rare-earth magnets. In fact most rare-earth magnets have not been discussed along the line of Kitaev interactions, except 
the first few works~\cite{PhysRevB.95.085132,Jang,RauYb}.
In the previous work~\cite{PhysRevB.95.085132}, we have illustrated this observation with the FCC rare-earth magnets. Since many non-honeycomb lattice iridates are claimed as Kitaev materials, it is thus reasonable to consider the rare-earth magnets with other crystal structures to be potential Kitaev materials beyond the previously proposed ones and the honeycomb one here~\cite{RauMichel,PhysRevLett.119.057203}. The reason that these rare-earth magnets contain a Kitaev interaction is due to two facts. The first fact is the spin-obital-entangled effective spin-1/2 local moment. The second fact is the 
three-fold rotation symmetry at the lattice site. 
This symmetry permutes the effective spin components and generates a Kitaev
interaction. These two ingredients can be used as the recipe to search for other rare-earth Kitaev materials beyond the honeycomb one. 
 
To summarize, we have focused on rare-earth honeycomb materials 
with nearest-neighbor interactions and computed the high-temperature 
thermodynamic properties, ESR linewidth, and spin-wave behaviors
as the experimental consequences of the anisotropic spin interaction.

\section{Acknowledgments}

We thank Leon Balents, Ni Ni, and Jiaqiang Yan for conversations. 
Gang Chen thanks Prof Peng Xue for her hospitality during the visit to 
Beijing Computational Science Research Center where this work is 
completed. This work is supported by the Ministry of Science and 
Technology of China with the Grant No.2016YFA0301001,2016YFA0300500,2018YFE0103200,
by from the Research Grants Council of Hong Kong with General 
Research Fund Grant No.17303819,
 by the Heising-Simons Foundation, 
and the National Science Foundation under Grant No. NSF PHY-1748958.

%\section*{Appendices}
\begin{appendix}

\section{Generic spin models and candidate states for higher spins}
\label{ssec1}

This spin model in Eq.~\eqref{eq:Hamiltonian} is designed for effective spin-1/2
local moments. It can be well extended to the high-spin local moments. 
For the honeycomb lattice with spin-1 local moments, the pairwise spin interaction is 
given as 
\begin{equation}
\begin{split}
H= & \sum_{\langle ij\rangle} J_{zz} S_i^z S_j^z + J_\pm (S_i^+S_j^- + S_i^-S_j^+)\\
& + J_{\pm\pm} (\gamma_{ij} S_i^+ S_j^+ +\gamma_{ij}^* S_i^- S_j^-)\\
& + J_{\pm z}[(\gamma_{ij}^* S_i^+ S_j^z+\gamma_{ij}S_i^-S_j^z)+\langle i \leftrightarrow j\rangle]\\
& + \sum_i D (S_i^z)^2.
\end{split}
\label{eqspin1}
\end{equation}
Because of the larger Hilbert space, a single-ion anisotropy is allowed and new states such
as the quantum paramagnet can be favored here. Thus, the phase transitions between quantum paramagnet
and other ordered phases can be interesting. Further neighbor exchange interaction, if included,
could bring more frustration channel than the spin-orbit entanglement induced frustration. It is 
known that, simple $J_1$-$J_2$ (first neighbor and second neighbor Heisenberg)
model on honeycomb lattice could induce spiral spin liquids in two dimensions where the spiral degeneracy has a line degeneracy in the momentum space rather than 
the surface degeneracy. The presence of the anisotropic interaction in Eq.~\eqref{eqspin1}
would overcome the quantum/classical order by disorder effect and lift the degeneracy. 
In addition to the spin-1 local moments, the model in Eq.~\eqref{eqspin1} also applies
to the spin-3/2 systems. Since the honeycomb lattice 
contains three nearest neighbor bonds, one may consider the possibility 
of the ALKT states on the honeycomb lattice where the nearest neighbor bonds 
are covered with spin singles of the spin-1/2 states and three onsite spin-1/2 spins are 
combined back to a spin-3/2 local moments.

Here, we have only listed the pairwise spin interactions. 
Due to the spin-orbital entanglement and the spin-lattice coupling,
the effective interaction for the spin-1 and spin-3/2 magnets 
can contain significant multipolar interactions. 
A simple example would be the biquadratic exchange 
$-({\boldsymbol S}_i \cdot {\boldsymbol S}_j)^2$
that is induced effectively by the spin-lattice coupling. 
 The presence of these 
multipolar interactions can significantly enhance quantum fluctuation
by allowing the system to tunnel more effectively within the 
local spin Hilbert space and thus create more quantum states such
as multipolar ordered phases and quantum spin liquids~\cite{PhysRevB.84.094420,PhysRevB.82.174440}.

The relevant physical systems for the spin-1 and spin-3/2 moments
would contain the $4d^2,5d^2,4d^4,5d^4$ and $4d^1,5d^1,4d^3,5d^3$  
magnetic ions, respectively. The relevant ions can arise from  
Ru, Mo and even V atoms, where spin-orbit coupling in the partially
filled $t_{2g}$ shell is active~\cite{PhysRevB.84.094420,PhysRevB.82.174440}.

\section{Details of high temperature expansion}
\label{ssec2}
	
The high temperature expansion requires to take into account the commutation relations between different spin operators on a same site. To this end, we define a vertex function $\nu_i (n_x, n_y, n_z)$ at each site $i$ following Ref.~\onlinecite{Fisher}, where $n_x, n_y$ and $n_z$ have to be even integers for the function to be nonzero. Its explicit form can be calculated by introducing the generating function
	\begin{equation}
	\psi(\xi, \eta, \zeta)=\text{Tr}~\left[ \exp (\xi S^x + \eta S^y + \zeta S^z) \right].
	\end{equation}
	Expanding the exponential and using the definition of $\nu$, we have
	\begin{equation}
	\psi(\xi, \eta, \zeta)=2 \sum_{n_x=0}^{\infty} \sum_{n_y=0}^{\infty} \sum_{n_z=0}^{\infty} \frac{\nu(n_x, n_y, n_z)}{2^{n_x+n_y+n_z}}  \frac{\xi^{n_x} \eta^{n_y} \zeta^{n_z}}{n_x! n_y! n_z!}.
	\end{equation}
	On the other hand, by diagonalizing the matrix of the exponential, we have
	\begin{equation}
		\psi(\xi, \eta, \zeta)=2\cosh\left( \sqrt{\xi^2+\eta^2+\zeta^2}/2 \right).
	\end{equation}
	Expanding this and comparing with the previous equation,
	\begin{equation}
		\nu(n_x, n_y, n_z)=\frac{[(n_x+n_y+n_z)/2]!}{(n_x/2)!(n_y/2)!(n_z/2)!}\frac{n_x! n_y! n_z!}{(n_x+n_y+n_z)!}.
	\end{equation}
We note this function is symmetric under the permutation of $n_x, n_y, n_z$.
%%%%%%%%%%%%%%%%%%%%%%%%%%%%%%%%%%%%%%%%%%%%
\begin{widetext}
The heat capacity is related to the zero-field partition function in the following way
\begin{equation}
C=\frac{1}{N}\frac{\partial E}{\partial T}=\frac{\beta^2}{N} \left[ \frac{1}{Z_0} \frac{\partial^2 Z_0}{\partial \beta^2}-\frac{1}{Z_0^2} \left(\frac{\partial Z_0}{\partial \beta}\right)^2\right],
\end{equation}
where we have divided by the number of sites $N$ to get the intensive quantity. $Z_0$ is given by
\begin{equation}
Z_0=2^N \left[1+ \frac{1}{4}\beta^2 \sum_{\langle i j\rangle}(\frac{1}{8}J_{zz}^2+J_\pm^2+J_{\pm\pm}^2+J_{\pm z}^2)\right]+O(\beta^3),
\end{equation}
where the $2^N$ factor results from the summation over all possible  configurations.

Susceptibility in direction $a$ can be reduced to the following expectation values of two spin operators,
\begin{equation}
    \chi_{a} =\frac{1}{\beta N} \frac{\partial^2}{\partial h_a^2} \ln Z \big|_{h_a=0}=\frac{\mu_0\mu_B^2g_a^2}{N Z_0}\beta \langle \sum_{m,n} S_m^a S_n^a\rangle_0.
\end{equation}
Using the vertex function $\nu(n_x, n_y, n_z)$ defined above, we obtain for the parallel case,
\begin{equation}
\begin{split}
& \langle \sum_{m,n} S_m^z S_n^z \rangle_{0}
= \sum_{\{\mathbf{S}_i\}} \sum_{m,n} S_m^z S_n^z e^{-\beta H}\\
= & \sum_{\{\mathbf{S}_i\}}\left[\sum_{m,n} S_m^z S_n^z -\beta \sum_{\langle i j\rangle}\sum_{m,n} H_{ij} S_m^z S_n^z+\frac{1}{2}\beta^2 \sum_{\langle i j\rangle}\sum_{\langle k l\rangle}\sum_{m,n} H_{ij} H_{kl} S_m^z S_n^z +\text{permutations} + O(\beta^3)\right]\\
= & \sum_{\{\mathbf{S}_i\}} \left\{ \frac{N}{4}-\frac{3N}{16}\beta J_{zz}+\frac{3N}{128}\beta^2 J_{zz}^2 +(\frac{3N^2}{32}-\frac{3N}{16})\beta^2(\frac{1}{8}J_{zz}^2+J_\pm^2+J_{\pm\pm}^2+J_{\pm z}^2) \right.\\
& + \frac{\beta^2}{32}(J_\pm^2+J_{\pm\pm}^2)\sum_{\langle i j\rangle}\left[\nu_i(2,0,2)\nu_j(2,0,0)+\nu_i(2,0,2)\nu_j(0,2,0)+\nu_i(0,2,2)\nu_j(2,0,0)+ \nu_i(0,2,2)\nu_j(0,2,0)\right]\\
& +\frac{\beta^2}{32}J_{\pm z}^2\sum_{\langle i j\rangle} \left[\nu_i(2,0,2)\nu_j(0,0,2)+\nu_i(0,2,2)\nu_j(0,0,2)+ \nu_i (0,0,4)\nu_j(2,0,0)+\nu_i(0,0,4)\nu_j(0,2,0)\right]\\
& \left.+\frac{1}{16}\beta^2 \sum_{\langle ij \rangle} (J_{zz}^2-2J_{\pm z}^2)+O(\beta^3)\right\}\\
=& 2^N\cdot \frac{N}{4}\left[ 1-\frac{3}{4}\beta J_{zz} + \beta^2( \frac{3}{8}J_{zz}^2-\frac{1}{2}J_\pm^2-\frac{1}{2}J_{\pm \pm}^2-J_{\pm z}^2) + \frac{3N}{8}\beta^2 (\frac{1}{8} J_{zz}^2+J_\pm^2+J_{\pm \pm}^2+J_{\pm z}^2)+O(\beta^3)\right].
\end{split}
\end{equation}
Here, the summation for $\{\mathbf{S}_i\}$ is over the possible configurations of spins on all sites. The notation $H_{ij}$ means the terms in the Hamiltonian for the bond labeled by sites $i, j$; namely, $H=\sum_{\langle i j\rangle}H_{ij}$.``Permutations'' on the second line are those with respect to the relative orderings of $H_{ij},$ $H_{kl},$ $S_m^z$ and $S_n^z$. 

Similarly, for the perpendicular susceptibility, we have 
\begin{equation}
\begin{split}
& \langle \sum_{m,n} S_m^x S_n^x \rangle_{0}
= \sum_{\{\mathbf{S}_i\}} \sum_{m,n} S_m^x S_n^x e^{-\beta H}\\
= & \sum_{\{\mathbf{S}_i\}}\left[\sum_{m,n} S_m^x S_n^x -\beta \sum_{\langle i j\rangle}\sum_{m,n} H_{ij} S_m^x S_n^x+\frac{1}{2}\beta^2 \sum_{\langle i j\rangle}\sum_{\langle k l\rangle}\sum_{m,n} H_{ij} H_{kl} S_m^x S_n^x +\text{permutations} + O(\beta^3)\right]\\
=& \sum_{\{\mathbf{S}_i\}} \left\{ \frac{N}{4}-\frac{3N}{8}\beta J_\pm + \frac{N-2}{16}\beta^2\sum_{\langle ij\rangle} (\frac{1}{8} J_{zz}^2 + J_{\pm}^2 + J_{\pm\pm}^2 + J_{\pm z}^2)+\frac{1}{64}\beta^2J_{zz}^2 \sum_{\langle ij \rangle} \nu_i(2,0,2)\nu_j(0,0,2)\right.\\
& +\frac{1}{32}\beta^2(J_\pm^2+J_{\pm\pm}^2)\sum_{\langle ij \rangle}[\nu_i (4,0,0)\nu_j(2,0,0)+\nu_i(4,0,0)\nu_j(0,2,0)+\nu_i(2,2,0)\nu_j(2,0,0)+\nu_i(2,2,0)\nu_j(0,2,0)]\\
& + \frac{1}{32}\beta^2 J_{\pm z}^2\sum_{\langle ij \rangle} [\nu_i (4,0,0)\nu_j(0,0,2)+\nu_i(2,2,0)\nu_j(0,0,2)+\nu_i(2,0,2)\nu_j(2,0,0)+\nu_i(2,0,2)\nu_j(0,2,0)]\\
\end{split}\nonumber
\end{equation}
\begin{equation}
\begin{split}
& \left. + \beta^2 \sum_{\langle ij \rangle} \sum_{\langle jk \rangle} \left[ \frac{1}{8} J_\pm^2 + \frac{1}{16} J_{\pm\pm}^2(\gamma_{ij}\gamma_{jk}^*+\gamma_{ij}^*\gamma_{jk}) + \frac{1}{32} J_{\pm z}^2 (\gamma_{ij}\gamma_{jk}+\gamma_{ij}\gamma_{jk}^*+c.c)\right]+O(\beta^3)\right\}\\
= & 2^N\cdot\frac{N}{4} \left[ 1- \frac{3}{2}\beta J_\pm+\frac{3N}{8}\beta^2 (\frac{1}{8} J_{zz}^2 + J_\pm^2+J_{\pm\pm}^2+J_{\pm z}^2)-\frac{1}{16}\beta^2 J_{zz}^2 + \frac{5}{4} \beta^2 J_\pm^2-\beta^2 J_{\pm\pm}^2 -\frac{3}{4} \beta^2 J_{\pm z}^2+O(\beta^3)\right].
\end{split}
\end{equation}

The magnetotropic coefficient $k$ can be computed using its relationship with the partition function with non-zero external field,
\begin{equation}
    k=\frac{1}{N}\frac{\partial^2 F}{\partial \theta^2}=\frac{1}{\beta N} \left[\frac{1}{Z^2} \left(\frac{\partial Z}{\partial \theta}\right)^2-\frac{1}{Z}\frac{\partial^2 Z}{\partial \theta^2}\right].
\end{equation}
The first term always give higher order terms compared with the second term, while the latter reads,
\begin{equation}
\begin{split}
    \frac{\partial^2 Z}{\partial \theta^2}= & -\beta \mu_0 \mu_B \sum_i \langle g_\perp \cos\theta (\cos\varphi S_i^x + \sin\varphi S_i^y) + g_\parallel \sin\theta S_i^z\rangle+\beta^2 \mu_0^2\mu_B^2 \sum_{i,j} \langle g_\perp^2 \sin^2\theta \cos^2\varphi S_i^x S_j^x + g_\perp^2\sin^2\theta \sin^2\varphi S_i^y S_j^y \\
& +g_\parallel^2\cos^2\theta S_i^z S_j^z-g_\perp g_\parallel \sin\theta\cos\theta\cos\varphi (S_i^z S_j^x+S_i^x S_j^z)-g_\perp g_\parallel\sin\theta\cos\theta\sin\varphi (S_i^z S_j^y+S_i^y S_j^z) \\
& +g_\perp^2 \sin^2\theta \sin\varphi\cos\varphi (S_i^x S_j^y+S_i^y S_j^x)\rangle+O(\beta^3)\\
= &  \mu_0^2\mu_B^2\beta^2\cos2\theta
(g_\parallel^2-g_\perp^2)+\frac{3\mu_0^2\mu_B^2}{4}\beta^3\cos2\theta(2g_\perp^2 J_\pm -g_\parallel^2 J_{zz})-\frac{\mu_0^4\mu_B^4}{48}\beta^4\cos4\theta (g_\perp^2-g_\parallel^2)^2(3N-2)\\
& -\frac{\mu_0^2\mu_B^2}{192}\beta^4 \cos2\theta \left\{ 3N(g_\perp^2-g_\parallel^2)\left[4\mu_0^2\mu_B^2(g_\perp^2+g_\parallel^2)+3(J_{zz}^2+8J_{\pm}^2+8J_{\pm\pm}^2+8J_{\pm z}^2)\right]\right.\\
& -\left.4\left[2\mu_0^2\mu_B^2(g_\perp^4-g_\parallel^4)-6g_\parallel^2(-3J_{zz}^2+4J_{\pm}^2+4J_{\pm\pm}^2+8J_{\pm z}^2)+3g_\perp^2(J_{zz}^2-20J_{\pm}^2+16J_{\pm\pm}^2+12J_{\pm z}^2)  \right]\right\}.
\end{split}
\end{equation}
The expression above reduces to, in the limit $g_\perp=g_\parallel$,
\begin{equation}
\frac{\partial^2Z}{\partial\theta^2}\big|_{g_\perp=g_\parallel}=2^N\cdot \frac{N}{4} \beta^3 \mu_0^2 \mu_B^2 \cos2\theta \left[-12 J_{zz}+24J_{\pm}+\beta(7J_{zz}^2-28J_{\pm}^2+8J_{\pm\pm}^2-4J_{\pm z}^2)\right].
\end{equation}
\end{widetext}
\end{appendix}

\bibliography{20200217}
\bibliographystyle{unsrt}

\end{document}